# Strain-induced gyrotropic effects in ferroelectric BaTiS$_3$

Wei Luo[1], Asier Zabalo[2], Guodong Ren[4], Gwan-Yeong Jung[5], Massimiliano Stengel[2,3], Rohan Mishra[4,5], Jayakanth Ravichandran[6,7] and Laurent Bellaiche[1,8]

[1]Smart Ferroic Materials Center, Physics Department and Institute for Nanoscience and Engineering, University of Arkansas Fayetteville, Arkansas 72701, USA

[2]Institut de Ciencia de Materials de Barcelona (ICMAB-CSIC), Campus UAB, 08193 Bellaterra, Spain

[3]ICREA-Institució Catalana de Recerca i Estudis Avançats, 08010 Barcelona, Spain

[4]Institute of Materials Science and Engineering, Washington University in St. Louis, St. Louis, MO 63130, USA

[5]Department of Mechanical Engineering and Material Science, Washington University in St. Louis, St. Louis, MO 63130, USA

[6]Mork Family Department of Chemical Engineering and Materials Science, University of Southern California, Los Angeles, CA 90089, USA

[7]Ming Hsieh Department of Electrical Engineering, University of Southern California, Los Angeles, CA 90089, USA

[8]Department of Materials Science and Engineering, Tel Aviv University, Ramat Aviv, Tel Aviv 6997801, Israel

Corresponding author: weil@uark.edu

**Abstract:**

Gyrotropic effects, including natural optical activity (NOA) and the nonlinear anomalous Hall effect (NAHE), are crucial for advancing optical and transport devices. We explore these effects in the $BaTiS_3$ system, a quasi-one-dimensional crystal that exhibits giant optical anisotropy. [Niu *et al*. Nat. Photonics 12, 392 (2018); Zhao *et al*. Chem. Mater. 34, 5680 (2022)]. In the *P6₃cm* phase which is stable under room temperature, we predict two distinct strain-induced phase transitions: a symmetry-lowering transition from the *P6₃cm* to *P6₃* phase under tensile strain (>3.0%), which enhances NOA and enables optical rotation; and an isostructural insulator-to-polar Weyl semimetal (WSM) transition under compressive strain (≤ –3%), which activates the NAHE and exhibits a strain-induced sign reversal. The low-temperature *P2₁* phase also transforms into a *P2₁2₁2₁* phase under enough compressive strains with such phase transition exhibiting a large NOA. All these results highlight $BaTiS_3$ as a viable candidate for novel ferroelectrics, optical and transport devices with strain enhanced or activated gyrotropic properties.

**Introduction**

Gyrotropic effects, including natural optical activity (NOA) and the nonlinear anomalous Hall effect (NAHE), play a crucial role in optical and transport devices. NOA is most commonly observed in chiral materials, such as quartz[1-3] and tellurium [4,5]. Nevertheless, chirality is not a strict requirement for NOA to occur, which has also been observed in achiral systems[6]. NAHE arises in metallic systems lacking inversion symmetry and has been widely studied both theoretically[7-16] and experimentally[17-19]. However, research on NAHE in polar Weyl semimetals (WSMs) remains scarce. This is primarily due to two challenges: (1) polar metals are rare because free electron screening suppresses spontaneous polarization, and (2) among the known polar metals, those hosting Weyl points at the Fermi level are even more scarce. Discovery of new polar WSMs can enable investigating the role of topology on NAHE, and help realize novel, nano-electronic transport devices based.

Recently, the hexagonal perovskite chalcogenide perovskite, BaTiS$_3$, has garnered significant attention due to its intriguing properties, including giant optical anisotropy[20,21], glass-like thermal conductivity[22,23] and multiferroicity[24]. Experimental studies have identified three distinct structural phases of BaTiS$_3$ across different temperature ranges[25]: a $P6_3cm$ state between 300 K and 240 K, a $P3c1$ phase between 240 K and ~150 K, and a $P2_1$ state below ~150 K[25,26]. Notably, the room-temperature $P6_3cm$ phase exhibits ferrielectric order[24], while the $P3c1$ phase is associated with a charge density wave (CDW) state[26]. The structural richness of BaTiS$_3$ motivates us to explore whether a transition between different phases can be achieved, allowing the diverse functionalities of BaTiS$_3$ to be utilized.

In this Letter, we conduct and analyze first-principles calculations to investigate the gyrotropic effects of both NOA and NAHE in the synthesized BaTiS$_3$[20,25,27] system for two different initial phases, namely the $P6_3cm$ phase experimentally found at room temperature and the $P2_1$ phase detected at low temperatures (<~150K). For each phase, we consider compressive and tensile biaxial in-plane strains. For the initial room-temperature phase under tensile strain (>3.0%), we numerically found a $P6_3cm$-to-$P6_3$ phase transition, as consistent with the work of Ref[24]. Such a transition increases the number of independent coefficients of the gyrotropic tensor from one to three, with some of these coefficients being significantly large at this transition and enabling optical rotation. Furthermore, under compressive strain smaller than -2%, we also predict an insulator-to-polar WSM topological phase transition within the same $P6_3cm$ space group, and which is accompanied by the occurrence of a spontaneous NAHE. Regarding the low-temperature phase, the magnitude of NOA is found to monotonously increase as the compressive strain decreases, and another transition (from $P2_1$-to-$P2_12_12_1$) occurs for compressive strains about -4.0%. The magnitude of NOA is found to be large near such latter transition. Our work therefore indicates that strained BaTiS$_3$ is a promising candidate for novel optical and transport devices.

**NOA for the room-temperature phase $P6_3cm$ with and without strain**

The crystal structure of the room-temperature $P6_3cm$ phase is shown in Fig. 1a. It can be viewed as a quasi-one-dimensional (1D) hexagonal chalcogenide with face-sharing $TiS_6$ octahedra chain along the $c$-axis. $P6_3cm$ $BaTiS_3$ has antipolar displacements of Ti atoms along the $c$-axis. It comprises three face-sharing $TiS_6$ octahedral chains; two chains show positive polarization (P), and the third one shows negative P, giving an overall ferrielectric order[24,28] (Fig. 1a). As indicated in Table S1 of the supporting information (SI), the lattice constants obtained in our first-principles calculations are slightly larger (by about 1.3%) than the experimental values, which is attributed to the overestimation of lattice constants by the Perdew–Burke–Ernzerhof (PBE) functional (see part I of the SI for details about these calculations). It is important to realize that (i) the point group of the $P6_3cm$ phase is $C_{6v}$ (6mm), allowing for an electric polarization along the z-direction, (ii) $P6_3cm$ is an achiral space group because it contains mirror symmetry; and (iii) from a symmetry perspective[29], the $P6_3cm$ space group permits only one independent gyrotropic tensor component $g_{12}$, as shown in the following matrix:

$$\begin{bmatrix} 0 & g_{12} & 0 \\ -g_{12} & 0 & 0 \\ 0 & 0 & 0 \end{bmatrix}$$

Interestingly, for tensile strain exceeding 3% and as shown in Fig. 2a, the ground state is not the $P6_3cm$ phase anymore but rather a $P6_3$ state. Technically, $P6_3$ belongs to the so-called Sohncke type of space groups[30], and this $P6_3$ phase is a chiral multiferroic phase with coupled ferroelectric and ferroaxial order[24] featuring quasi-1D $TiS_6$ octahedral chains that rotate either in an anti-clockwise (ACW) or clockwise (CW) direction (see Figs. 1b and 1c). In addition, the Ti atoms exhibit polar displacements along the $c$-axis of the chains, with neighboring chains having parallel alignment of the dipoles, resulting in a finite polarization. In this $P6_3$ phase, three independent gyrotropic tensor components are allowed, and the corresponding matrix[29] is given by:

$$\begin{bmatrix} g_{11} & g_{12} & 0 \\ -g_{12} & g_{11} & 0 \\ 0 & 0 & g_{33} \end{bmatrix}$$

We computed the gyrotropic tensor for both $P6_3cm$ and $P6_3$ phase (ACW) of BaTiS$_3$ under different strains using density-functional perturbation theory (DFPT), as implemented in the *Abinit* code[31-33] (see part I of SI for details). The results are shown in Fig. 2b. Note that the misfit strain $\eta_{misfit}$ ranges approximately between -6% to 6% and is defined as:

$$\eta_{misfit} = \frac{a-a_0}{a_0}, \tag{1}$$

where $a$ represents the in-plane lattice constant of the substrate, while $a_0$ is the in-plane lattice constant obtained from the energy minimization of the investigated structure. To model perfect epitaxy on a cubic substrate, the strain tensor, expressed in Voigt notation, has three *fixed* components during each simulation, which are:

$$\eta_1 = \eta_2 = \eta_{misfit}, \text{ and } \eta_6 = 0. \tag{2}$$

On the other hand, $\eta_3$, $\eta_4$, and $\eta_5$ are allowed to relax, along with all internal atomic coordinates. For $\eta_{misfit} \leq 3\%$, only $g_{12}$ is nonzero since the phase remains $P6_3cm$. Additionally, $g_{12}$ increases as the strain decreases down to the compressive value of -2%. Strikingly and as detailed in Fig. S1 of the SI, an insulator-to-metal phase transition occurs when the compressive strain is larger (in magnitude) than -2% (note that therefore we only present the gyration tensor for strains larger than -2%, since the DFPT calculation of the gyration tensor is only applicable to insulating systems). Furthermore, the gyration tensor coefficient $g_{12}$ exhibits a discontinuity, and the $g_{11}$ and $g_{33}$ suddenly become finite (with $g_{33}$ being rather big) at the phase transition from $P6_3cm$ to $P6_3$. For $\eta_{misfit} > 4\%$, these gyration tensor components $g_{11}$, $g_{12}$ and $g_{33}$ all decrease monotonically as the tensile strain increases.

We now discuss how optical rotation—i.e., circular birefringence, the rotation of the polarization plane of linearly polarized light as it passes through an optically active medium—can be experimentally observed in room-temperature BaTiS$_3$. Optical rotation is typically observed in uniaxial crystals when light propagates along the optic axis, since linear birefringence vanishes in that direction. In other directions, linear birefringence dominates, making the detection of optical rotation challenging, as its magnitude is much smaller than that of linear birefringence. Both the $P6_3cm$ and $P6_3$ phases of BaTiS$_3$ are uniaxial crystals, with their optical axis aligned along the $c$-direction. It is important to note that the refractive index behavior is determined solely by the symmetric part of the gyrotropic tensor[34]. Therefore, when considering the experimental observation of optical rotation, only the symmetric component of the gyrotropic tensor is relevant. For completeness, and to support potential future studies, we provide both the symmetric and antisymmetric components of the gyrotropic tensor in our calculations. A detailed discussion on the role of the antisymmetric terms can be found in Ref.[34].

In the $P6_3cm$ phase, the gyrotropic tensor contains only an antisymmetric component; therefore, optical rotation cannot be observed. In contrast, for the $P6_3$ phase (ACW), the symmetric part of the gyrotropic tensor is given as follows:

$$\begin{bmatrix} g_{11} & 0 & 0 \\ 0 & g_{11} & 0 \\ 0 & 0 & g_{33} \end{bmatrix}$$

The symmetric part of the gyrotropic tensor in the $P6_3$ phase has therefore two independent components ($g_{11}$ and $g_{33}$). To understand how the optical rotation power is related to the gyrotropic tensor in this phase, we derived the optical equation from Maxwell's equations (The detailed derivation can be found in Part II of the SI)

$$(\sum_j s_i s_j)E_j + \sum_j[\varepsilon_{ij} + i\eta_{ijm}G_m]E_j = n^2 I E_i. \tag{3}$$

The $\boldsymbol{s}$ elements are related to the propagation vector, and $\boldsymbol{E}$ is the electric field of the light. $\varepsilon$ is the relative dielectric permittivity. $\eta_{ijm}$ is the permutation tensor, which satisfies $\eta_{123} = -\eta_{213} = 1$ and $\eta_{113} = 0$. $G$ is the gyrotropic vector. $n$ denotes the refractive index corresponding to light propagation along $\boldsymbol{s}$. $I$ is the 3×3 identity matrix. By solving this optical equation, we find that the optical rotation is governed by the $g_{33}$ component of the gyrotropic tensor and is given by:

$$\rho = \frac{\pi(n_L - n_R)}{\lambda} = \frac{\omega^2}{2c^2} g_{33}. \tag{4}$$

Here, $\rho$ is the optical rotation power and $\lambda$ is the wavelength of light. The $n_L$ and $n_R$ denote the refractive indices for left- and right-circularly polarized light, respectively. The $\omega$ and $c$ are circular frequency of incident light and the speed of light, respectively. In the ACW state of the $P6_3$ phase where the polarization is along the positive $c$-axis, a linearly polarized light beam traveling along the negative $c$-direction (see Fig. 1b) will undergo a clockwise rotation upon passing through the sample. Interestingly too, in the CW of the $P6_3$ phase, the same linearly polarized light will be rotated counterclockwise (see Fig. 1c), as the gyrotropic tensor changes sign with the reversal of crystal chirality. This finding indicates that optical rotation can be controlled by ferroelectric polarization—a property not found in classical optically active materials such as $\alpha$-quartz or trigonal selenium. By comparing the $P6_3cm$ and $P6_3$ phases, we observe that optical rotation emerges as a result of the structural phase transition, which is induced by tensile strain—since the $P6_3cm$ phase does not exhibit optical rotation. Conversely, optical rotation can serve as a probe for detecting the structural phase transition[24], as it is zero in the $P6_3cm$ phase and non-zero in the $P6_3$ phase.

**NAHE for the room-temperature $P6_3cm$ phase of BaTiS$_3$ under strain**

As indicated above and as evidenced by the band structure with and without spin-orbit coupling (SOC) calculations (see part I, Figs. S1 and S2 of SI for details), an insulator-to-polar WSM phase transition takes place within the $P6_3cm$ state at a compressive strain near -3%. Moreover, Fig. 3a depicts the corresponding band structure of the $P6_3cm$ phase with SOC at -4% strain. One can clearly observe the aforementioned insulator-to-WSM phase transition. The polar WSM phase of $P6_3cm$ BaTiS$_3$ is unusual since most polar phase are insulators. We hypothesize that the stability of the polar metal phase in BaTiS$_3$ arises from the weak screening effects of Weyl electrons. As a matter of fact, since WSMs have Weyl nodes near the Fermi level, the electronic density of states remains low, leading to inherently weak electron screening.

Interestingly, we find that the Weyl points are located precisely at the Fermi level, indicating that the system is an ideal Weyl semimetal[35,36]. Notably, other strain conditions—namely −3.0%, −5.0%, and −6.0%—also exhibit ideal Weyl semimetal behavior, as their Weyl points likewise lie at the Fermi level (see Fig. S2 in the SI). More intriguingly, we identify this Weyl semimetal as a Berry dipole semimetal[37] (also referred to as a dipolar Weyl semimetal[38]), characterized by a pair of Weyl points in close momentum proximity (detailed in Table S2 of the SI). The Berry curvature texture under −4.0% strain is shown in Fig. 3b and is clearly distinguishable from that of an isolated Weyl point[39].

Due to the breaking of inversion symmetry, the NAHE emerges, with the magnitude of the nonlinear anomalous Hall conductivity being proportional to the Berry curvature dipole[7] (BCD). For the $P6_3cm$ phase, which belongs to the $C_{6v}$ point group, the non-zero components of the BCD are determined by the following symmetry constraints:

$$D = \det(S)\, SDS^{-1}. \tag{5}$$

Here, $S$ denotes the orthogonal symmetry operation matrix, and $D$ represents the Berry curvature dipole (BCD) tensor. The determinant of the matrix, $\det(S)$, equals 1 for rotational symmetries and −1 for mirror or inversion symmetries. Our analysis reveals that the BCD tensor possesses only one independent non-zero component, namely $D_{xy}$, and that the BCD retains only its antisymmetric part, transforming as a pseudovector under symmetry operations. (Details on the derivation of the non-zero components of the BCD tensor are provided in part III of the SI.) The BCD tensor takes the following form:

$$\begin{bmatrix} 0 & D_{xy} & 0 \\ -D_{xy} & 0 & 0 \\ 0 & 0 & 0 \end{bmatrix}$$

Here, we calculate the BCD (see part I of SI for details) as a function of chemical potential for compressively strained $BaTiS_3$ at -3%, -4%, -5%, and -6% strains – that is within the polar WSM phase. The results are presented in Figs. 3c, 3d, 3e, and 3f. Indeed, our results show that $D_{xy}$ is non-zero and satisfies the antisymmetric relation $D_{xy} = -D_{yx}$, in agreement with the predictions from our symmetry analysis.

The BCD ($D_{xy}$) exhibits a peak near the Fermi level under −3% (−0.083 eV), −4% (−0.100 eV), −5% (−0.104 eV), and −6% (−0.081 eV) strains, as indicated by the black solid lines in Figs. 3c–3f. As the strain decreases in magnitude from −4% to −3%, the BCD ($D_{xy}$) undergoes a notable sign reversal (see green curves in Figs. 3c and 3d). To understand the origin of this reversal, we consider the definition of BCD, $D_{xy} = \int_{BZ}[d\mathbf{k}] \sum_n \frac{\partial E_{kn}}{\partial k_x} \Omega^y_{kn}(-\frac{\partial f_0}{\partial E})_{E=E_{kn}}$. It is evident from this expression that the sign of the BCD is determined by the product of the band velocity $\frac{\partial E_{kn}}{\partial k_x}$ (i.e., the slope of the band along the $k_x$ direction) and the Berry curvature $\Omega^y_{kn}$. To examine their behavior, we plot the band structures and the Berry curvature component $\Omega_y$ along the Γ–X path for −3% and −4% strain in Figs. 4c and 4d, respectively. While the band slopes along Γ–X remain negative in both cases, the Berry curvature profiles differ significantly. For −4% strain, $\Omega_y$ remains consistently negative along the Γ–X direction. In contrast, for −3% strain, $\Omega_y$ exhibits strong oscillations and integrates to a net positive value over the same path. This difference in the Berry curvature distribution is thus responsible for the sign reversal of the BCD between −4% and −3% strain. Moreover, we hypothesize that this qualitative change in $\Omega_y$ arises from the strain-induced structural modification. Figure 4a shows the $c/a$ ratio as a function of strain from 0% to −6%. The data can be fitted by two distinct linear trends: one for the 0% to −3% range and another for −4% to −6%, with differing slopes. This discontinuity in the slope indicates an abrupt change in the $c/a$ ratio between −3% and −4%, which may underlie the transformation in the Berry curvature texture and, consequently, the sign reversal in the BCD.

From −4.0% to −6.0% strain, the sign of the BCD $D_{xy}$ remains unchanged. However, its magnitude at the energy peak ($E_p$) gradually decreases as the compressive strain increases in magnitude (see Figs. 3d–3f). This reduction is attributed to the corresponding decrease in the magnitude of the Berry curvature component $\Omega_y$ over the same strain range (Figs. 4d–4f). We hypothesize that the diminishing $\Omega_y$ is correlated with a weakening of inversion symmetry breaking under increasing compressive strain. To test this hypothesis, we calculated the strength of such breaking, via the relative

displacements (along the $c$-direction) of Ti atoms between the FE and paraelectric (PE) states:

$$\Delta d_z = \sum_{i=1,6}[d_z(FE) - d_z(PE)]_i. \tag{6}$$

Here, $i$ runs over the six Ti atoms in the unit cell, and $d_z(FE)$ and $d_z(PE)$ are the displacements of the Ti atoms along the $c$-direction in the FE and PE states, respectively. Note that, as indicated above, the $P6_3cm$ phase has the $C_{6v}$ point group, which permits polarization along the $c$-direction – which explains why we focused on the Ti displacements along the $z$-axis. The calculated $\Delta d_z$ of Eq. (6) for different strains is shown in Fig. 4b. It is clear that $\Delta d_z$ is large for the 0% strain case and decreases as the strain becomes more compressive. To explain why the magnitude of $\Omega_y$ decreases from −4% to −6% strain, we propose that this is linked to a reduction in the strength of inversion symmetry breaking, $\Delta d_z$ (see Fig. 4b), between −4% and −6% strain. In contrast, from −3% to −4%, $D_{xy}$ increases while $\Delta d_z$ decreases, which appears to contradict the expected trend. This discrepancy can be attributed to the large strain effect between −3% and −4%, which significantly alters the characteristics of $\Omega_y$ (compare Figs. 4c, 4d, 4e, and 4f). Consequently, the trend of decreasing $D_{xy}$ from −4% to −6% strain holds, as the distribution of $\Omega_y$ remains similar, with only the magnitude differing.

These results therefore reveal that the BCD in $BaTiS_3$ is highly tunable, as its magnitude and sign is very sensitive to strain. It is also noteworthy that other gyrotropic properties, like kinetic magnetic effect (KME)[40] should also exist in strained $BaTiS_3$ ($P6_3cm$) for compressive strains greater than -2%. Here, we want to emphasize that the decrease in $\Delta d_z$ with increasing the magnitude of the compressive strain is unusual, as the atomic displacement in the FE state typically increases with an increase in the $c/a$ ratio[41,42] (or compressive strain). This atypical behavior is likely related to the electronic properties of the system, specifically the insulator-to-polar metal phase transition. As compressive strain increases in magnitude, this phase transition occurs, and mobile electrons screen the ionic polarization (i.e., the ionic displacement). Consequently, the ionic displacement $\Delta d_z$ decreases as compressive strain increases.

**NOA of the low-temperature $P2_1$ phase of BaTiS$_3$ under strain**

Around 100 K, the ground state of BaTiS$_3$ is the $P2_1$ phase. $P2_1$ phase has antipolar displacement of Ti atoms along the $c$-axis. It comprises four TiS$_6$ octahedral chains – two with positive P and two with negative P, leading to an overall antiferroelectric order, as illustrated in Fig. 1d. Under compressive strain smaller than -4%, we found that the $P2_1$ phase evolves into a $P2_12_12_1$ phase, which also exhibits antiferroelectric order (see Fig. 1e). The gyrotropic tensor matrix for the $P2_1$ phase is

$$\begin{bmatrix} g_{11} & g_{12} & 0 \\ g_{21} & g_{22} & 0 \\ 0 & 0 & g_{33} \end{bmatrix}$$

And thus have five independent values. In contrast, for the $P2_12_12_1$ phase, there are only three independent values, and the gyrotropic tensor matrix is given by:

$$\begin{bmatrix} g_{11} & 0 & 0 \\ 0 & g_{22} & 0 \\ 0 & 0 & g_{33} \end{bmatrix}$$

The calculated gyrotropic tensors for the $P2_1$ and $P2_12_12_1$ phases are shown in Fig. 2c. It can be observed that, prior to the aforementioned phase transition, as compressive strain increases in magnitude, the magnitude of the gyrotropic tensor coefficients for the $P2_1$ phase all increase first weakly and then significantly near the transition. For instance, for $P2_1$ with a strain around -4%, $g_{22}$ is notably large in magnitude, approximately -6.1 Bohr. For the $P2_12_12_1$ phase, as compressive strain increases in magnitude, the magnitudes of $g_{11}$, $g_{22}$ and $g_{33}$ increase slightly.

Note that we also calculated the gyrotropic tensor components $g_{11}$, $g_{22}$ and $g_{33}$ for compressive strain, but excluding the local-field self-consistent fields (SCFs), and the results are shown in Table I. It is evident that the local-field SCFs play a significant role in determining the magnitude and strain behavior of the gyrotropic tensor, as consistent with previous studies[31].

**Conclusion**

In conclusion, our first-principles study of the ferroelectric (antiferroelectric for $P2_1$ and $P2_12_12_1$) BaTiS$_3$ system has revealed significant strain-induced modifications of its gyrotropic properties in both the initial $P6_3cm$ room-temperature and $P2_1$ low-temperature phases, including the enhancement of NOA, the facilitation of optical rotation, the activation of NAHE, and the strain-induced sign reversal of NAHE. Such features are found to be related to various factors, such as (i) the existence of strain-induced $P6_3cm$-to-$P6_3$ and $P2_1$-to-$P2_12_12_1$ phase transitions; (ii) the occurrence of an isostructural transition from an insulator to a polar WSM phase within the $P6_3cm$ state; and (iii) modification of the strength of inversion symmetry breaking with strains. All these results demonstrate the tunability of gyrotropic effects through strain engineering and provide new insights into the potential of BaTiS$_3$ as a candidate material for novel optical and transport devices.

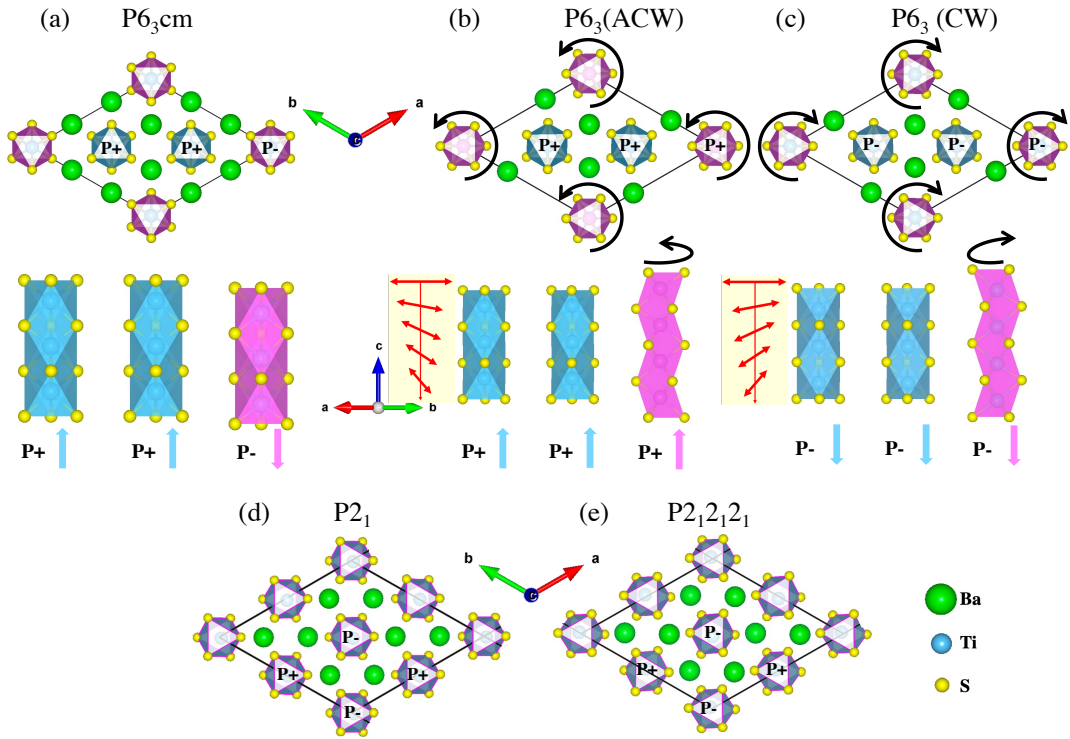

Fig. 1: Top views of the crystal structure of BaTiS$_3$: (a) room-temperature $P6_3cm$ phase; (b) $P6_3$ phase with anti-clockwise Ti-S 1D chain rotation; (c) $P6_3$ phase with clockwise chain rotation; (d) low-temperature $P2_1$ phase; (e) low-temperature $P2_12_12_1$ phase.

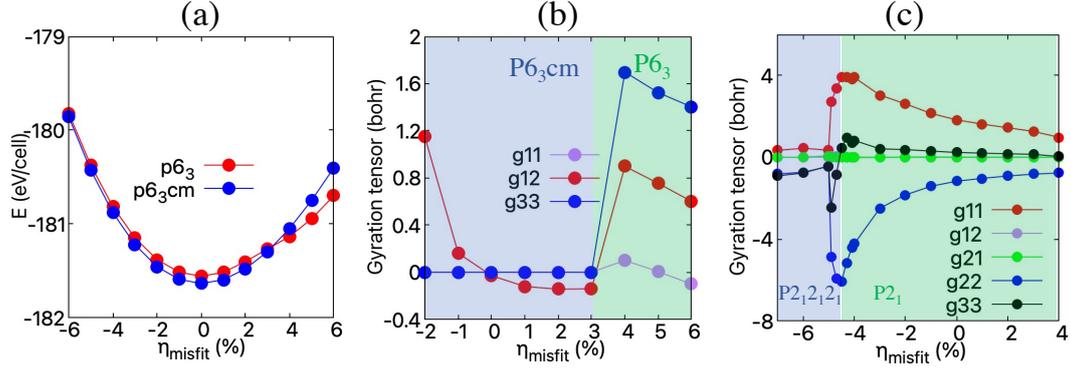

Fig. 2: (a) DFT total energies of the $P6_3$ and $P6_3cm$ phases as a function of strain. (b) Gyrotropic tensor coefficients vs. strain for the $P6_3cm$ (blue) and $P6_3$ (green) phases. (c) Same for the $P2_12_12_1$ (blue) and $P2_1$ (green) phases.

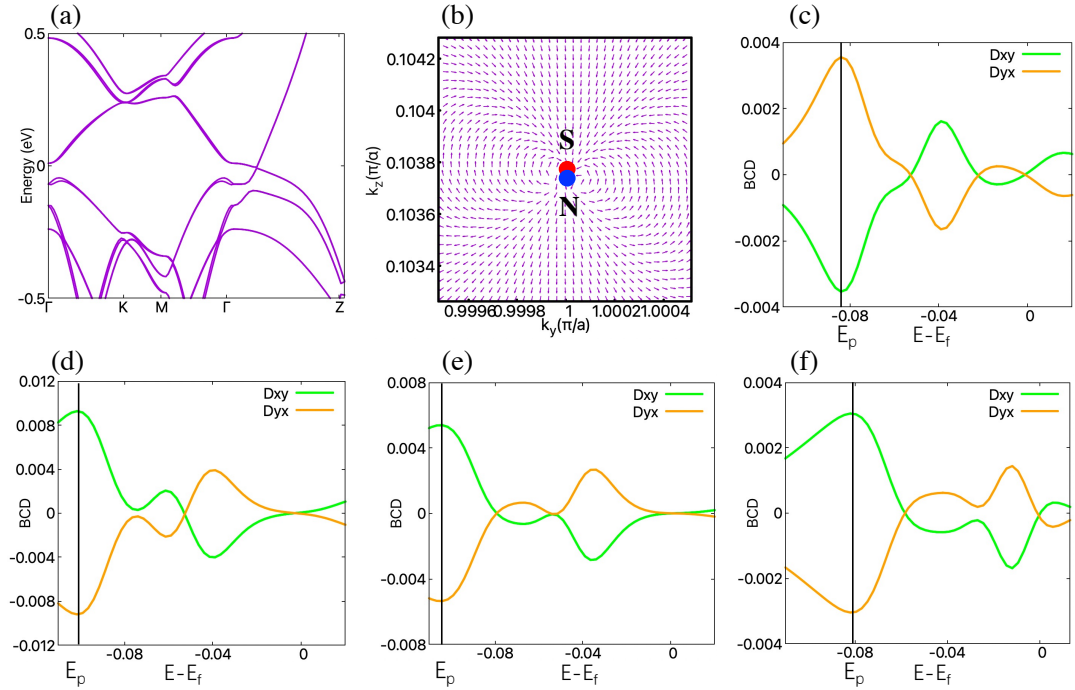

Fig. 3: (a) Band structure with SOC for the -4% strain. (b) Berry curvature texture of the Berry dipole semimetal, with S and N denoting the south and north Berry curvature monopoles, respectively. The BCD components $D_{xy}$ ($D_{yx}$) under $-3\%$ (c), $-4\%$ (d), $-5\%$ (e), and $-6\%$ (f) strain.

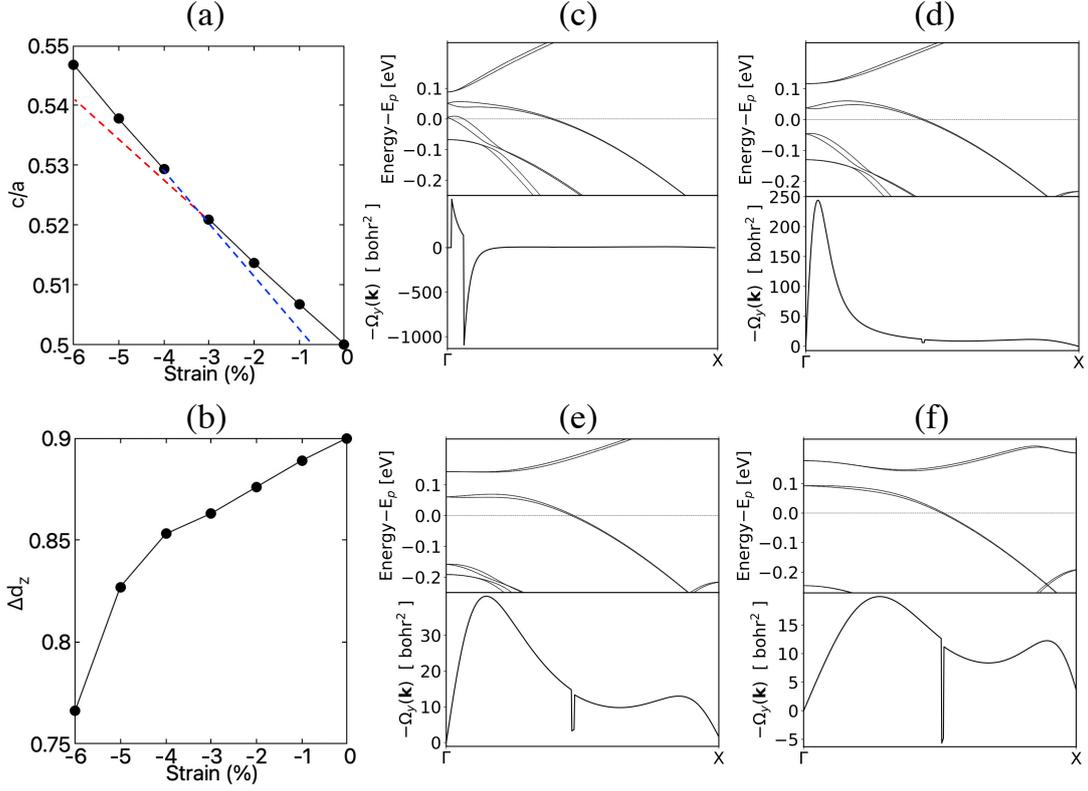

Fig. 4: (a) The $c/a$ ratio as a function of in-plane biaxial strain. The red dashed line is the straight line fitting the data from 0% to -3%, while the blue dashed line is the straight line fitting the data from -4% to -6%. (b) $\Delta d_z$ as a function of strain. Band structures and Berry curvature along $\Gamma$–X under –3% (c), –4% (d), –5% (e), and –6% (f) strain. $E_p$ denotes the chemical potential indicated in Figs. 3c–f.

Table I: Calculated gyrotropic tensor components $g_{11}$, $g_{22}$ and $g_{33}$ under –5% ($P2_12_12_1$), –4% ($P2_1$) and –3% ($P2_1$) strain. Values in parentheses exclude SCF terms.

|  | $g_{11}$ | $g_{22}$ | $g_{33}$ |
|---|---|---|---|
| -5% ($P2_12_12_1$) | 0.34 (0.26) | -0.45 (-0.25) | -0.43 (-0.27) |
| -4% ($P2_1$) | 3.90 (2.11) | -4.23 (-2.17) | 0.82 (1.2) |
| -3% ($P2_1$) | 3.03 (1.55) | -2.49 (-1.35) | 0.41 (0.08) |

**Acknowledgments**

We thank Dr. L. Y. Gao for insightful discussions. This research is supported by the Arkansas High Performance Computing Center which is funded through multiple National Science Foundation grants and the Arkansas Economic Development

Commission. W. L and L. B. acknowledge the support from the Vannevar Bush Faculty Fellowship (VBFF) Grant No. N00014-20-1-2834 from the Department of Defense, the ARO Grant No. W911NF-21-1-0113, and the MonArk NSF Quantum Foundry supported by the National Science Foundation Q-AMASE-i Program under NSF Award No. DMR-1906383. A. Z and M. S. acknowledge support from Ministerio de Ciencia e Innovación (MICINN-Spain) through Grant No. PID2019– 108573 GB-C22; from Severo Ochoa FUNFUTURE center of excellence (CEX2019-000917-S); from Generalitat de Catalunya (Grant No. 2021 SGR 01519); and from the European Research Council (ERC) under the European Union's Horizon 2020 research and innovation program (Grant Agreement No. 724529). G. R., G. J. and R. M. acknowledge support from the National Science Foundation (NSF) through grant numbers DMR-2145797 and DMR-2122070. J. R. acknowledge the support from the ARO MURI program with award number W911NF-21-1-0327, and the National Science Foundation (NSF) through grant number: DMR-2122071.